\documentclass[aps,prd,twocolumn,superscriptaddress,amsmath,amssymb]{revtex4-1}
\usepackage{graphicx} 
\usepackage{epstopdf} 
\usepackage{dcolumn} 
\usepackage{xcolor} 
\usepackage{amssymb} 
\usepackage{hyperref} 
\usepackage[utf8]{inputenc} 
\usepackage[english]{babel} 
\usepackage[displaymath, mathlines]{lineno} 
\usepackage{blindtext} 
\usepackage{comment}
\usepackage{amsmath,bm}
\usepackage{subfigure,subfigmat}
\usepackage{orcidlink}

\graphicspath{{figures/}} 

\usepackage{enumitem}  
\usepackage{feynmf}    
\newcommand{\overbar}[1]{\mkern 1.5mu\overline{\mkern-1.5mu#1\mkern-0.1mu}\mkern 0.1mu}
\input belle2-symbols.tex

\begin{document}


\title{Measurement of \cp asymmetries in \ksksks decays at Belle~II
}
  \author{I.~Adachi\,\orcidlink{0000-0003-2287-0173}} 
  \author{L.~Aggarwal\,\orcidlink{0000-0002-0909-7537}} 
  \author{H.~Ahmed\,\orcidlink{0000-0003-3976-7498}} 
  \author{H.~Aihara\,\orcidlink{0000-0002-1907-5964}} 
  \author{N.~Akopov\,\orcidlink{0000-0002-4425-2096}} 
  \author{A.~Aloisio\,\orcidlink{0000-0002-3883-6693}} 
  \author{N.~Anh~Ky\,\orcidlink{0000-0003-0471-197X}} 
  \author{D.~M.~Asner\,\orcidlink{0000-0002-1586-5790}} 
  \author{H.~Atmacan\,\orcidlink{0000-0003-2435-501X}} 
  \author{T.~Aushev\,\orcidlink{0000-0002-6347-7055}} 
% ...
\collaboration{The Belle II Collaboration}

\begin{abstract}
We report a measurement of decay-time dependent charge-parity (\cp) asymmetries in \ksksks decays.
We use $387 \times 10^6$ \BB pairs collected at the \FourS resonance with the Belle II detector at the SuperKEKB asymmetric-energy electron-positron collider. 
We reconstruct 220 signal events and extract the \cp-violating parameters \scp and \ccp from a fit to the distribution of the decay-time difference between the two $B$ mesons. 
The resulting confidence region is consistent with previous measurements in \ksksks and $\Bz \ra (c\overbar{c})K^0$ decays, and with predictions based on the standard model.

\end{abstract}

\maketitle

\section{Introduction}
In the standard model (SM), the charmless three-body decay \ksksks is mediated by the $b \to sq\overbar{q}$ quark transition, which is dominated by a one-loop process, the so-called penguin amplitude. 
Charge-conjugate decays are implied hereafter unless specified otherwise.
Penguin amplitudes are suppressed in the SM, e.g, ${\mathcal B}(B^0 \to K^0_S K^0_S K^0_S) = (6.0 \pm 0.5) \times 10^{-6}$~\cite{Workman:2022ynf}, and imply exchanges of virtual particles where SM particles can be replaced by a broad class of non-SM particles. These features make these decays sensitive to possible contributions from non-SM physics~\cite{Grossman:1996ke}. A key probe of such contributions is provided by decay-time dependent \cp-violating asymmetries of the $B^0$ and \Bzb decay rates. These asymmetries arise from interference between amplitudes for direct decay and decay following flavor oscillations, due to the irreducible phase in the Cabibbo-Kobayashi-Maskawa (CKM) quark-mixing matrix~\cite{Kobayashi:1973fv}. Precise measurements of these asymmetries using $\Bz\Bzb$ pairs are a primary goal of experiments in electron-positron collisions at the \FourS resonance. If one of the neutral $B$ mesons,  $B_{\cp}$, decays into a \cp eigenstate $f_{\cp}$ at proper time $t_{\cp}$ and the other, $B_{\rm tag}$, decays into a flavor-specific final state $f_{\rm tag}$ at proper time $t_{\rm tag}$, the probability density for observing a $B_{\rm tag}$ with flavor \qf at  $\Delta t \equiv t_{\cp} - t_{\rm tag}$ is~\cite{Carter:1980hr,Carter:1980tk,Bigi:1981qs}

\begin{linenomath}
\begin{align}\label{eqn_dt1}
\begin{split}
\mathcal{P}(\Delta t,\qf) =   \frac{e^{-|\Delta t|/\tau_{B^0}}}{4\tau_{B^0}} \biggl(1 + \qf\bigl[ \scp\sin(\Delta m^{}_d\Delta t) \\
- \ccp\cos(\Delta m^{}_d\Delta t)  \bigr] \biggr),
\end{split}
\end{align}
\end{linenomath}
where the flavor \qf is $+1 (-1)$ for $B_{\rm tag}=\Bz(\Bzb)$, $\tau_{B^0}$ is the $B^0$ lifetime, $\Delta m_d$ is the mass difference between the two mass eigenstates of the $B^0$-$\overbar{B}^0$ system, and the \cp asymmetries \scp ~and \ccp ~express mixing-induced and direct \cp violation, respectively\footnote{The coefficients $(S,C)$ are written $(S, -A)$ elsewhere.}.
The SM predicts that $\scp = -\sin 2 \phi_1 - 0.02$ and $\ccp = -0.007$ for decays into the \cp-even final state $\KS\KS\KS$~\cite{Cheng:2005ug}. The mixing phase  $\phi_1 \equiv \arg [-V_{cd}V_{cb}^*/V_{td}V_{tb}^*]$ is a combination of CKM matrix-elements. The uncertainty in the SM prediction for \scp is smaller than 0.01; hence, a large deviation in \ksksks decays would indicate non-SM physics. The Belle~\cite{Belle:2020cio} and BaBar~\cite{BaBar:2011ktx} experiments reported these asymmetries with comparable uncertainties dominated by the sample size, yielding world-average values $\scp = -0.83\pm0.17$ and $\ccp = -0.15\pm0.12$~\cite{PhysRevD.107.052008}. While these agree with the SM predictions, the large uncertainties limit the sensitivity to non-SM sources. Additional measurements are needed.

We report a measurement of \scp and \ccp in \ksksks decays using electron-positron collisions at the \FourS collected by the Belle II experiment.
We reconstruct signal (\BCP) \ksksks decays followed by $\KS \ra \pipi$ decays and suppress background using two multivariate classifiers. We then measure \qf using the remaining charged particles in the event and \deltat from the distance between the decay positions of \BCP and \Btag. We divide the \ksksks events into two classes based on the quality of the \deltat information: time-differential (TD) events use \deltat and determine \scp and \ccp, while time-integrated (TI) events do not use \deltat and contribute to the determination of \ccp only.  Fits to signal-discriminating observables and decay time (when appropriate) determine the signal yield and \cp asymmetries.
We use the decay \ksksk as a control channel to constrain the fit model from data.

\section{The Belle II detector and data sample}
The Belle~II experiment is located at SuperKEKB,
which collides electrons and positrons at and near the $\Upsilon(4S)$
resonance~\cite{Akai:2018mbz}. The Belle II
detector~\cite{Abe:2010gxa} has a cylindrical geometry and includes a six-layer silicon detector (VXD) and a 56-layer central drift chamber~(CDC). These detectors reconstruct trajectories of charged particles (tracks).  
The VXD consists of two layers of silicon-pixel detectors~(PXD) surrounded by four layers of double-sided silicon-strip detectors~\cite{Belle-IISVD:2022upf}.
Only the innermost PXD layer, and one sixth of the outermost layer are installed for the data analyzed here.
The symmetry axis of these detectors, defined as the $z$ axis, is almost
coincident with the direction of the electron beam.  Surrounding the
CDC, which also provides $dE/dx$ energy-loss measurements, is a
time-of-propagation counter~\cite{Kotchetkov:2018qzw} in the
central region and an aerogel-based ring-imaging Cherenkov
counter in the forward region.  These detectors provide
charged-particle identification.  Surrounding them is an
electromagnetic calorimeter based on CsI(Tl) crystals that
primarily provides energy and timing measurements for photons and
electrons. Outside of the calorimeter is a superconducting solenoid
magnet. The magnet provides a 1.5~T magnetic field parallel to the $z$ axis. Its flux return is instrumented with resistive-plate chambers
and plastic-scintillator modules to detect muons, $K^0_L$ mesons, and
neutrons.

We use data collected at the \FourS resonance in 2019--2022, corresponding to an integrated luminosity of \luminosity~and containing \nBBs pairs. We use simulated samples to train the multivariate classifiers and define fit models.
The $e^+e^- \ra \FourS \ra B\overbar{B}$ sample is generated using {\sc evtgen}~\cite{Lange:2001uf} and {\sc pythia}~\cite{Sjostrand:2014zea}. In the simulated signal sample, one of the $B$ mesons decays to the \ksksks signal mode or the \ksksk control mode according to phase space. The simulated $e^+e^- \ra q\overbar{q}$ sample, where $q$ indicates an $u$, $d$, $s$, or $c$ quark, is generated using the {\sc kkmc}~\cite{Jadach:1999vf} generator interfaced with {\sc pythia}. We also use {\sc evtgen} to simulate the decay of short-lived particles.
The detector response is simulated by {\sc geant4}~\cite{Agostinelli:2002hh}.
Experimental and simulated data are analyzed with the Belle II software~\cite{Kuhr:2018lps,basf2-zenodo}.

\section{Event reconstruction}
The $\Upsilon (4S)$ is produced at the $e^+e^-$ collision point with a Lorentz boost (${\beta \gamma}$) of 0.288
and subsequently decays to a $B$ and a $\overbar{B}$ meson, which are both nearly at rest in the $e^+e^-$ center-of-mass (c.m.)\ frame.
Therefore, the $B$-meson pairs propagate nearly along the boost direction with known velocity in the laboratory. This allows one to approximate the difference between their decay times as $\deltat = (\lboost - \tagvlboost)/\beta\gamma c$, where $z_{\cp({\rm tag})}$ is the decay position of $B_{\cp({\rm tag})}$ projected onto the boost axis.

Events are selected online based on the number of charged particles and total energy deposited in the calorimeter with nearly 100\% efficiency.
Pairs of oppositely-charged particles are used to reconstruct $\KS \to \pi^+ \pi^-$ candidates. 
The four-momentum and decay vertex of the \KS candidate are obtained from a kinematic fit of the $\pi^+$ and $\pi^-$ tracks. To reduce combinatorial background from incorrectly reconstructed \KS candidates, we use a boosted-decision-tree (BDT) classifier \fbdtks with 15 input variables that include the \KS flight length, the impact parameters of the \KS candidate and the $\pi^{\pm}$,
and the number of measurement points (hits) in the VXD associated with the $\pi^{\pm}$.
The most discriminating variables are the angle between the \KS momentum and the displacement of the \KS decay vertex from the beam interaction point (IP) and the \KS flight length normalized by its uncertainty.
We select \KS candidates with invariant mass $M(\pi^+\pi^-)$ between 462.6\mevcc and 532.6\mevcc, corresponding to about 35 units of the relevant resolution, and with an \fbdtks requirement that accepts 91\% of \KS mesons. The mass window is wide since the BDT efficiently suppresses the background. These criteria are optimized as described later.

We reconstruct \BCP candidates by combining three \KS candidates and treat the particles not belonging to \BCP as \Btag decay products.
We select \BCP candidates using the invariant mass \mkkk and the beam energy constrained mass $\mbc \equiv \sqrt{E_{\rm beam}^2 - |\vec{p}_{B}|^2c^2}/c^2$, where $E_{\rm beam}$ and $\vec{p}_B$ are the beam energy and the momentum of the $B$ meson in the $e^+e^-$ c.m.\ frame. 
We retain \BCP candidates satisfying $5.2 < \mbc < 5.29 \gevcc$ and $5.08 < \mkkk < 5.48 \gevcc$, but exclude those satisfying $5.265 < \mbc < 5.29 \gevcc$ and $5.08 < \mkkk < 5.2 \gevcc$ to avoid contamination by $B^{0(+)} \ra \KS \KS K^{*0(+)}$ decays.

The dominant source of background is the $e^+ e^- \to q\overbar{q}$ continuum.
We suppress this background by using another BDT classifier, \fbdtcs, with the following input variables that exploit event topology:
the cosine of the angle between the thrust axes of \BCP and \Btag in the $e^+e^-$ c.m.\ frame; the magnitude of the  \Btag thrust; the sum of the transverse momenta of the particles in the event; the squared four-momentum difference between the beams and the detected particles in the c.m.\ frame; and the modified Fox-Wolfram moments~\cite{Belle:2003fgr}.
The $B$ thrust axis is a unit vector $\hat{t}$ that maximizes the thrust magnitude $T \equiv \left(\sum_{i}\left|\hat{t}\cdot\vec{p}_i\right|\right)/\left(\sum_{i}\left|\vec{p}_i\right|\right)$, where $\vec{p}_i$ is the momentum of the $B$ meson's $i$-th decay-product  in the c.m.\ frame. The BDT classifier \fbdtcs ranges from zero for background-like events to one for signal-like events. We use simulated events to train the classifier. A requirement of $\fbdtcs > 0.1$ results in 51\% background rejection with a signal efficiency of 98\%. We then calculate a transformation of the classifier, $\modcs = \log\left[(\fbdtcs-0.1)/(1-\fbdtcs)\right]$, which yields a classifier distribution more convenient to parametrize.
The selection criteria on \KS candidate mass and \fbdtks are determined by maximizing $N_{\rm sig}/\sqrt{N_{\rm sig}+N_{\rm bkg}}$, where $N_{\rm sig}$ and $N_{\rm bkg}$ are the expected yields of \BCP signal and background events determined from simulation, respectively, meeting the following signal-enhancement conditions: $5.27 < \mbc < 5.29 \gevcc$, $5.18 < \mkkk < 5.38 \gevcc$, and $\fbdtcs > 0.5$.

In addition to the nonresonant decay amplitude, quasi-two-body decays $\Bz \ra X (\ra \KS \KS) \KS$ via intermediate resonances $X$ due to $b\ra s$ and $b \ra c$ transitions contribute to \ksksks decays. We consider $b\ra s$ decays to be signal, but we veto $b\ra c$ contributions to measure the \cp asymmetries for the $b \ra s$ transition.
We expect a significant $b \ra c$ contribution only from $\Bz \ra \chi_{c0} \KS$ decays based on the rates of $\Bz \ra X(\ra \KS \KS) \KS$ decays where $X$ indicates a $\overline{D}{}^0, \chi_{c0}, \chi_{c1}, \chi_{c2}, \eta_c, J/\psi,$ or $\psi(2S)$ meson~\cite{Workman:2022ynf}. The $\Bz \ra \chi_{c0} (\ra \KS \KS) \KS$ branching fraction is around 5\% of the signal branching fraction.
We reject signal \BCP candidates if the invariant mass of any combination of two \KS candidates is in the range $3.379 < M(\KS\KS) < 3.447 \gevcc$. This requirement rejects 90\% of the background from $\Bz \ra \chi_{c0} (\ra \KS \KS) \KS$ decays and 7.2\% of signal.

The control channel \ksksk is reconstructed from two \KS mesons and a track and is similar to the signal decay. We require the particle identification information for the track to be consistent with a $K^+$.
We use the control channel to constrain the parameters of $B$-vertex-resolution model for signal, as well as those of the shapes of the \BCP mass and \modcs background distributions.
We do not veto $\chi_{c0} K^+$ decays for the control channel because their kinematic distributions are the same as those of the $\KS \KS K^+$ final state.

\section{Measurement of $B$-meson flavor and decay-time difference}
We use a category-based BDT algorithm to identify the \Btag flavor~\cite{Belle-II:2021zvj}. The algorithm uses 13 BDTs, each geared toward discriminating a specific signature of $b \ra c \ra s$ cascade decays using particle identification and kinematic variables of the \Btag charged decay products.
The outputs from these BDTs are combined by the top-level BDT to return the flavor value \qf and tagging quality $r \equiv 1- 2w$, where $w$ is the probability for wrong flavor assignment. The probability density of Eq.~\eqref{eqn_dt1} is modified to include the parameter $w$, and its difference between $B^0$ and $\overbar{B}^0$, $\Delta w$,
\begin{linenomath}
\begin{align}\label{eqn_dt3} 
\begin{split}
\mathcal{P}^{\rm TD}_{\rm sig}&(\Delta t,\qf) =  \frac{e^{-|\Delta t|/\tau_{B^0}}}{4\tau_{B^0}}  
 \Big(\mathstrut^{\mathstrut}_{\mathstrut} 1 - \qf\Delta w \\ 
&  + \qf(1-2w) [ \scp\sin(\Delta m^{}_d\Delta t) - \ccp\cos(\Delta m^{}_d\Delta t) ] \Big) . 
\end{split}
\end{align}
\end{linenomath}
The events are classified into seven independent $r$ intervals (bins).
For each bin, $w$ and $\Delta w$ are determined using flavor-specific $B$ meson decays with large branching fractions~\cite{flavortag}. 
Since the signal purity varies as a function of $r$, using the distribution of $r$ improves the statistical sensitivity to the \cp asymmetries.

To measure \deltat, we reconstruct the \BCP and \Btag decay vertices
using information about the IP. The spatial distribution of the IP is described by a three-dimensional Gaussian whose parameters are regularly measured in a calibration based on $e^+e^- \ra \mu^+\mu^-$ events. The IP size is typically 250\mum in the boost direction, 10\mum in the horizontal direction, and 0.3\mum in the vertical direction~\cite{Belle-II:2022plj}.
The \BCP vertex position is reconstructed from the six final-state pions using a decay-chain vertex fit, which constrains the \BCP to originate from the IP (IP constraint)~\cite{TreeFit}.
Due to their long lifetime, a fraction of \KS mesons decay outside of the VXD volume resulting in poorly measured decay positions. This causes the \BCP vertex resolution to depend strongly on the number of \KS mesons with associated VXD hits.
In simulation, the fractions of signal decays in which zero, one, two, or three \KS mesons have VXD hits are 0.4\%, 7.9\%, 37.9\%, or 53.8\%, respectively. When only one \KS meson has VXD hits, the IP constraint significantly improves the \BCP vertex resolution, reducing the average vertex-position uncertainty in the boost direction from around 270\mum to 120\mum.
The average uncertainty with the IP constraint is 49\mum when two \KS mesons have VXD hits and 35\mum when all three have such hits.

We use the \Btag tracks to reconstruct the \Btag vertex, excluding those having no associated PXD hits. We also exclude pairs of oppositely-charged pions consistent with a \KS decay because they are likely to be produced away from the \Btag vertex.
Similarly to the \BCP vertex, we constrain the \Btag to originate from the IP to improve the vertex resolution and reconstruction efficiency~\cite{Dey:2020dsr}.
In order to reduce the contamination from tracks from secondary and tertiary displaced vertices, which would bias the determination of the \BCP vertex position, the fit is repeated by iteratively removing the tracks contributing the largest increase to the vertex-fit \chisq until a satisfactory fit quality is achieved. A selection on fit quality and vertex-position uncertainty is applied to ensure the quality of the \deltat measurement. 

We divide the remaining \ksksks candidates into two classes based on the quality of the \deltat information to maximize the sensitivity of the measurement of \scp and \ccp. 
For the time-differential (TD) analysis that determines both \scp and \ccp, we require candidates that satisfy the following criteria: both tracks from one or more signal \KS are associated with at least one VXD hit, the decay-time-difference satisfies $-30 < \deltat < 30\ps$, satisfactory vertex-fit quality, and small vertex-position uncertainty.
The \deltat resolution is around 0.9\ps in the TD events.
The \deltat information of the other events is not used. They are included in the time-integrated (TI) analysis, which contributes only to \ccp.
The probability density in Eq.~\eqref{eqn_dt3} is integrated over \deltat for TI events, yielding
\begin{linenomath}
\begin{align}\label{eq:Psig_timeint}
\mathcal{P}_{\rm sig}^{\rm TI}(\qf) = \frac{1}{2} \left( 1 - \qf \Delta w - \qf (1 - 2w) \ccp\frac{1}{1 + \Delta m_d^2 \tauBz^2} \right).
\end{align}
\end{linenomath}

For 1.1\% of simulated signal events, multiple (typically two) \BCP candidates are reconstructed. We choose the candidate with the best vertex-fit quality for such events, which retains the correctly reconstructed \BCP candidates in 82\% of these events. This requirement has negligible impact on the \deltat distribution and the \cp asymmetry results. The reconstruction efficiency including the \Btag selection is 28.3\% in simulation.
For the control channel, we reconstruct the \ksksk vertex without using the $K^+$ track to emulate the \ksksks vertex fit. We discard \ksksk candidates that fail the TD criteria. The reconstruction efficiency for the control channel is 24.7\%.

\section{Determination of signal yield}
We extract the yields for TD, TI, and control channel events from a three-dimensional likelihood fit to the unbinned distributions of \mbc; \mkkh, where $K$ indicates a \KS or $K^+$ meson; and \modcs.
The likelihood function includes two sample components, signal and background.
We determine the shape of the signal component from fits to distributions of simulated signal and control samples.
The \mbc distribution is modeled with a Gaussian function for the signal TD and control samples and with a Crystal Ball shape~\cite{Gaiser:Phd,Skwarnicki:1986xj} for the signal TI sample. The signal and control-sample \mkkh distribution is modeled with the sum of a Gaussian function and an asymmetric Breit-Wigner function. The signal and control-sample \modcs distribution is modeled with the sum of a symmetric and an asymmetric Gaussian function. 
For the background, the \mbc distribution is modeled with an ARGUS function~\cite{ARGUS:1990hfq}, the \mkkh distribution with a linear function, and the \modcs distribution with the sum of a symmetric and an asymmetric Gaussian function. 
The endpoint of the ARGUS function is set to $E_{\rm beam}$, which is calibrated using other $B$ decays.
The parameter sets for the \modcs shapes are shared between TD and TI events. 
We use the same parameter set for the \mbc and \mkkh background shapes across the three samples as the \ksksk kinematic properties are similar to those of the signal decay, as confirmed in simulation. The fit simultaneously determines the yield of each sample and 14 background shape parameters~\cite{MINUIT}.

\begin{figure}
    \includegraphics[width=0.9\columnwidth]{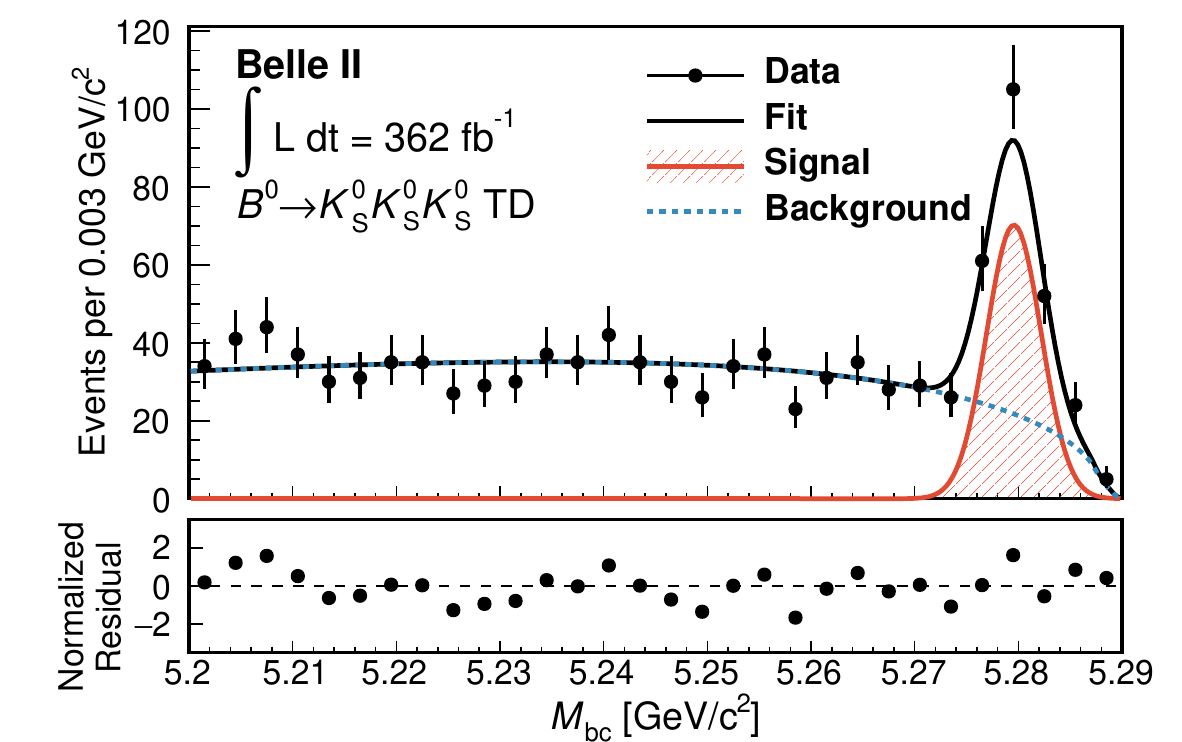}
    \includegraphics[width=0.9\columnwidth]{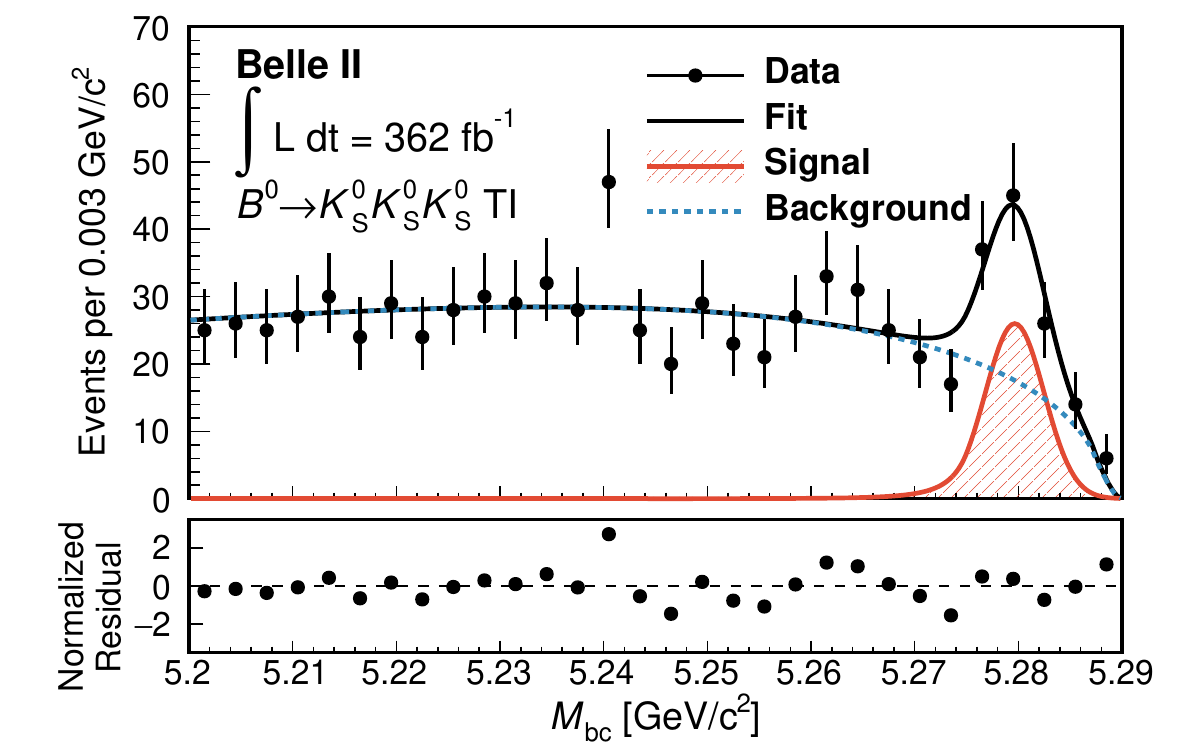}
    \includegraphics[width=0.9\columnwidth]{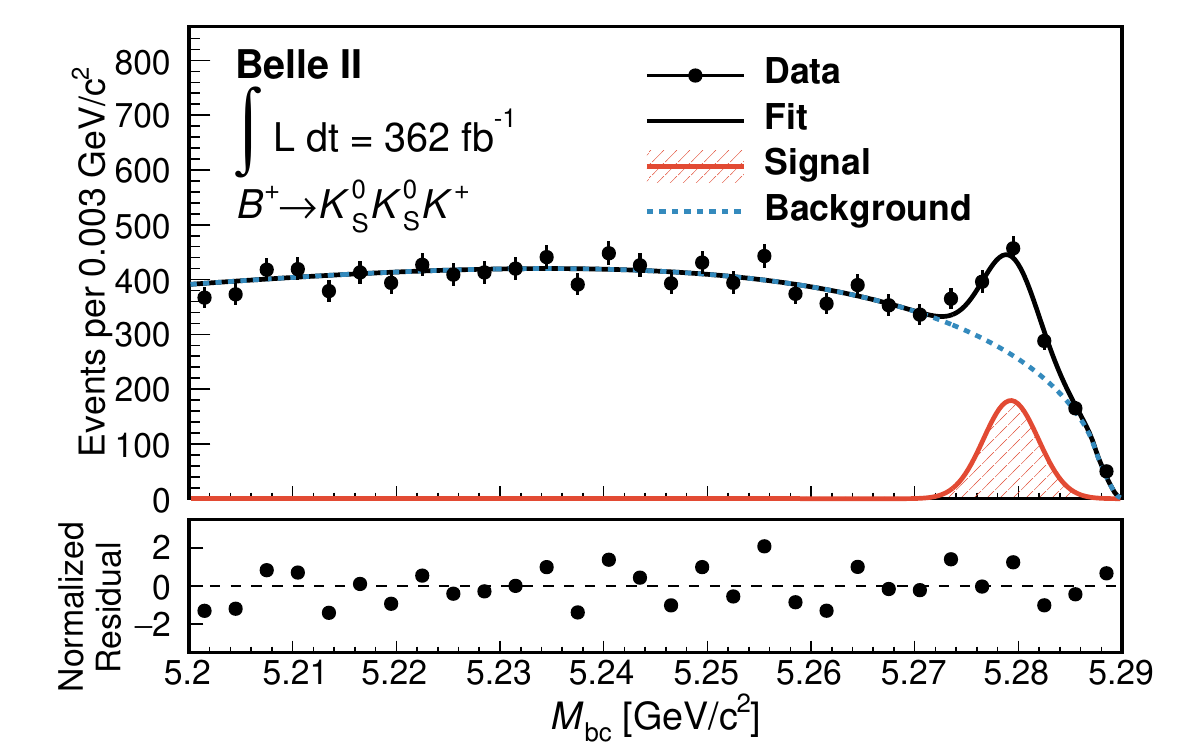}
  \caption{Distributions of \mbc for (top) TD, (middle) TI, and (bottom) \ksksk candidates with fit projections overlaid.
The black dots with error bars represent the data points; the black, solid curve shows the total fit projection; the red hatched area is the signal projection; and the blue, dashed curve is the background projection.
The distributions are restricted to events in the \mkkh signal region.
Lower panels show the differences between data and fit results normalized by the statistical uncertainty of the data.
}
\label{fig:mbc}
\end{figure}

\begin{figure}
    \includegraphics[width=0.9\columnwidth]{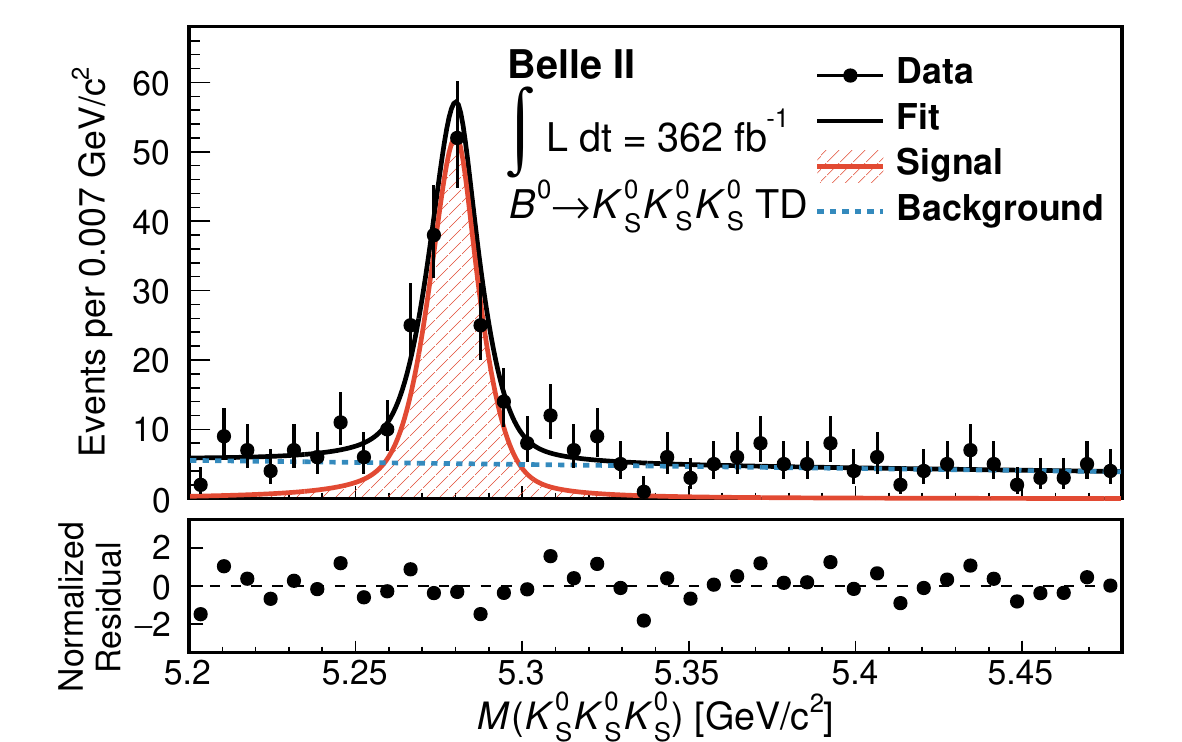}
    \includegraphics[width=0.9\columnwidth]{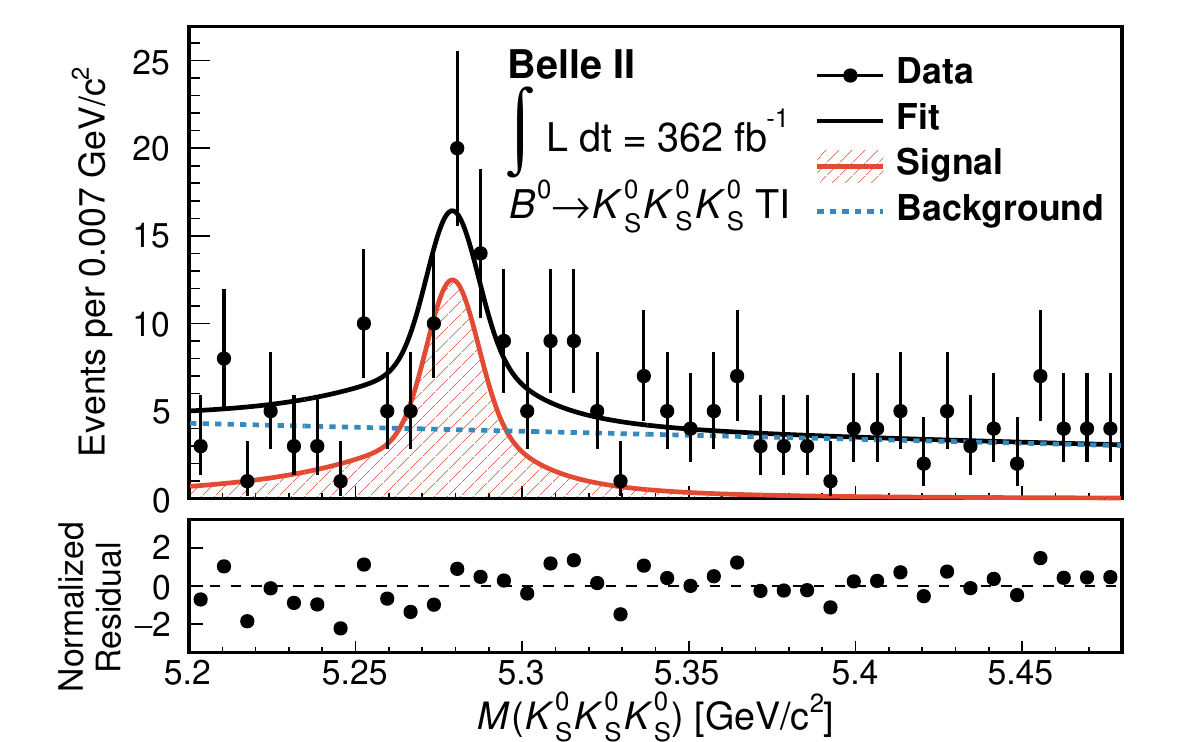}
    \includegraphics[width=0.9\columnwidth]{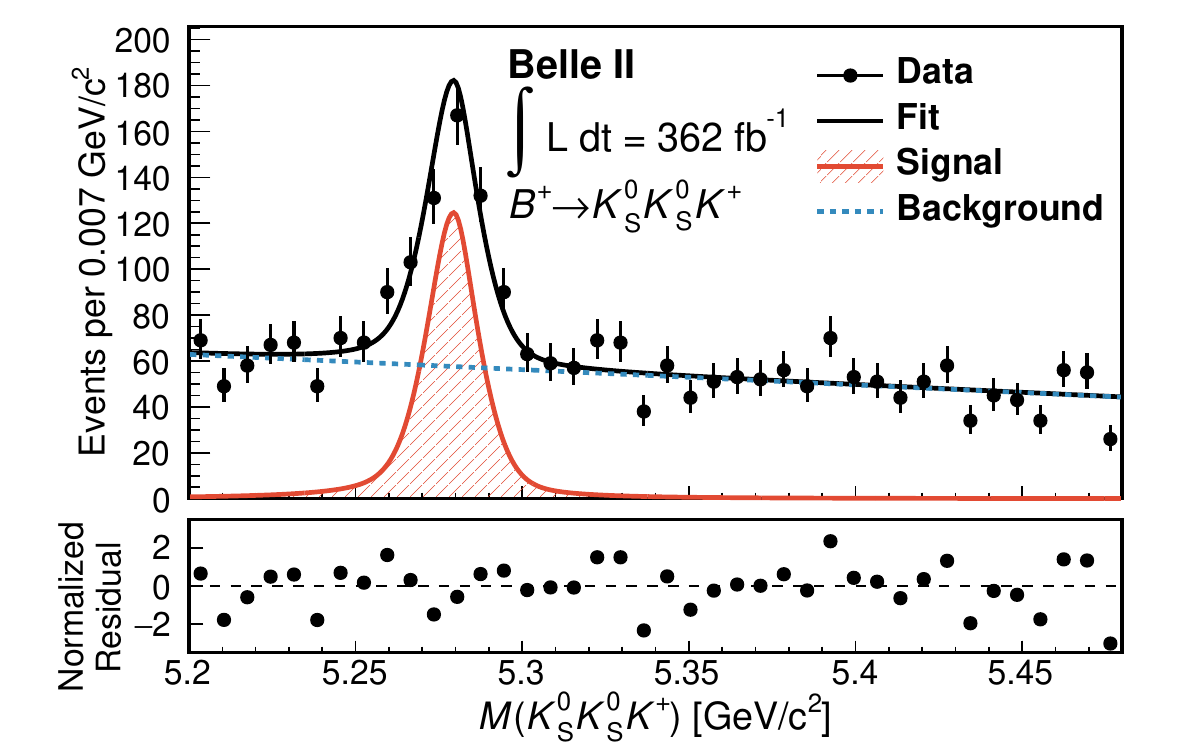}
  \caption{Distributions of \mkkh for (top) TD, (middle) TI, and (bottom) \ksksk candidates with fit projections overlaid.
The black dots with error bars represent the data points; the black, solid curve shows the total fit projection; the red hatched area is the signal projection; and the blue, dashed curve is the background projection.
The distributions are restricted to events in the \mbc signal region.
Lower panels show the differences between data and fit results normalized by the statistical uncertainty of the data.
}
\label{fig:mkkh}
\end{figure}

\begin{figure}
    \includegraphics[width=0.9\columnwidth]{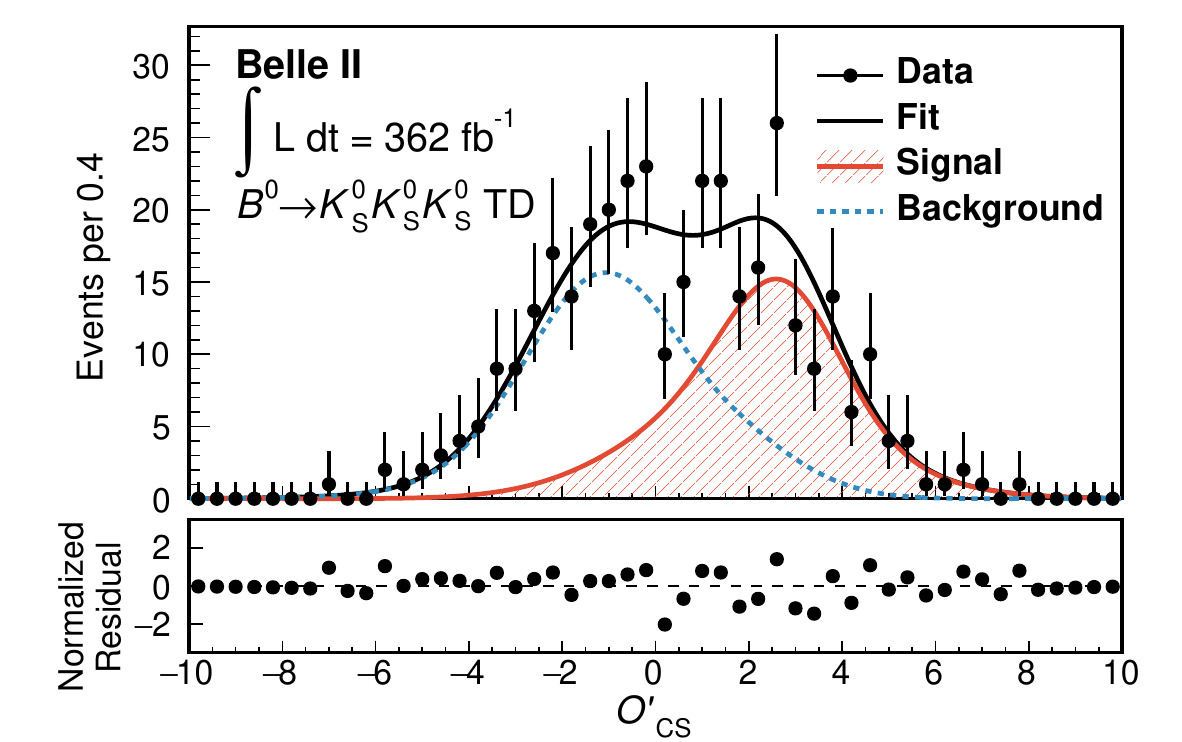}
    \includegraphics[width=0.9\columnwidth]{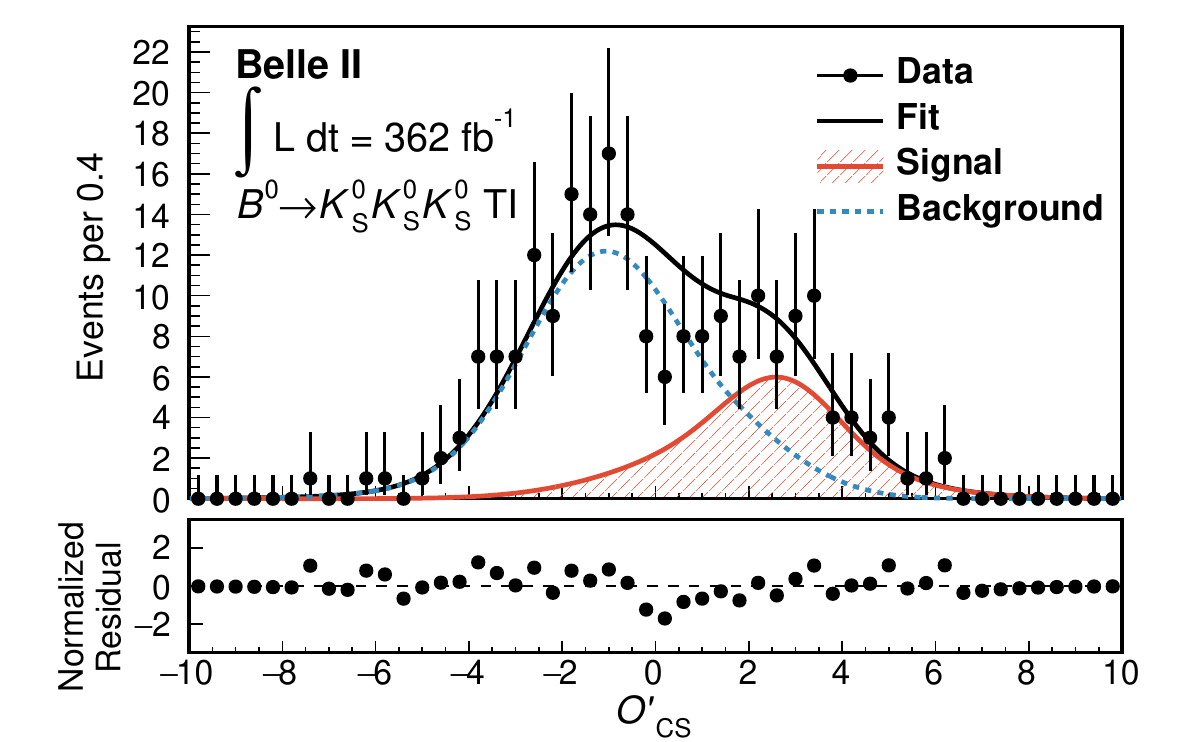}
    \includegraphics[width=0.9\columnwidth]{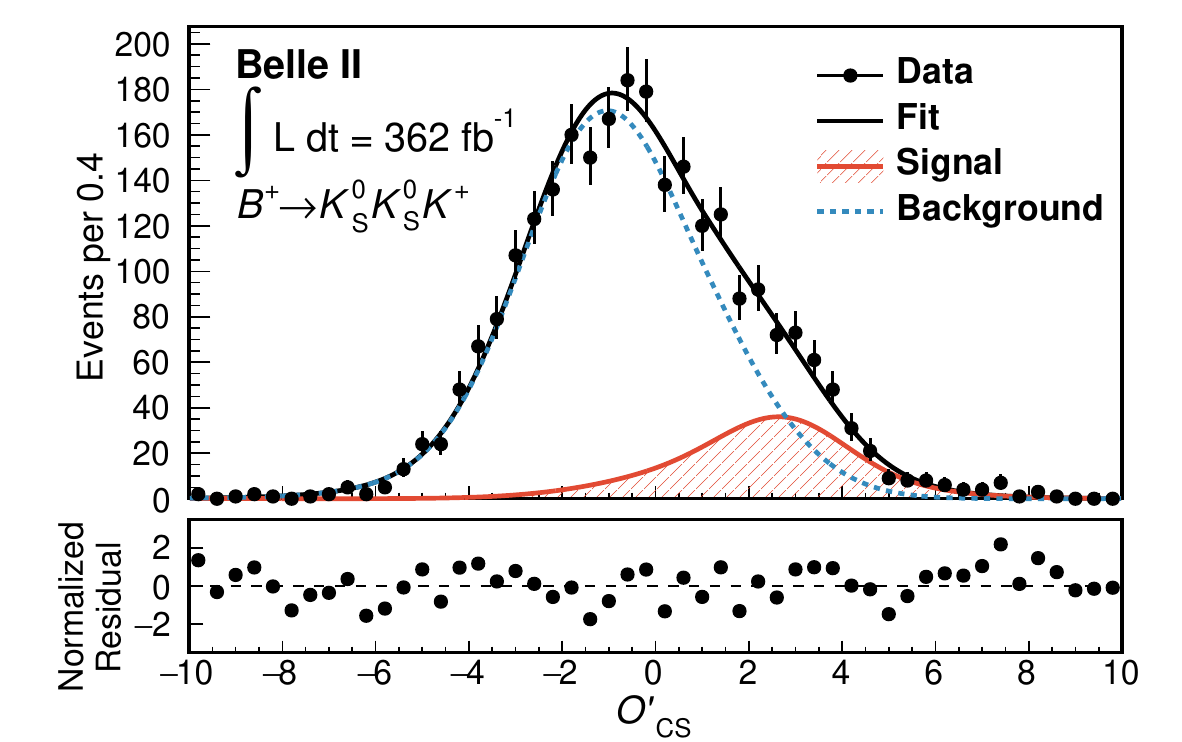}
  \caption{Distributions of \modcs for (top) TD, (middle) TI, and (bottom) \ksksk candidates with fit projections overlaid.
The black dots with error bars represent the data points; the black, solid curve shows the total fit projection; the red hatched area is the signal projection; and the blue, dashed curve is the background projection.
The distributions are restricted to events in the \mbc signal region.
Lower panels show the differences between data and fit results normalized by the statistical uncertainty of the data.
}
\label{fig:cs}
\end{figure}

Figures~\ref{fig:mbc}, \ref{fig:mkkh}, and \ref{fig:cs} show the data distributions with fit results overlaid. The low-mass tails of the \mkkh distribution of the signal TI component is mainly due to $\pi \ra \mu \nu_{\mu}$ decays, which occur in 3\% of the reconstructed signal \ksksks events. 
Such events are mostly classified as TI events due to the poor vertex fit quality.
We define the signal region as $5.272 < \mbc < 5.288 \gevcc$, $5.2 < \mkkh < 5.36\gevcc$, and $-4.44 < \modcs < 8.85$. 
Each range for \mbc and \modcs retains 99.73\% of signal TD events. 
The signal yield and the purity in the signal region is $158 ^{+14}_{-13}$ and 57\% for TD events, $62\pm9$ and 40\% for TI events, and $403^{+24}_{-23}$ and 22\% for the control channel events.

\section{Determination of \cp asymmetries}
We determine the \cp asymmetries \scp~and \ccp~from a maximum-likelihood fit to the unbinned \deltat and binned \qf distributions combining TD, TI, and \ksksk events restricted to the signal region.
The contribution to the likelihood function from the $i$-th TD event is 
\begin{linenomath}
\begin{align}\label{eqn_dt4}
\begin{split}
\mathcal{L}_{i}^{\rm TD}&(\scp, \ccp|\deltat_i, \qfi) = \\ 
&f^{\rm sig}_i \int d(\Delta t') R(\Delta t_i - \Delta t') \mathcal{P}^{\rm TD}_{\rm sig} (\Delta t',\qfi) \\
& + (1-f^{\rm sig}_{i}) \mathcal{P}_{\rm bkg} (\Delta t_i),
\end{split}
\end{align}
\end{linenomath}
where $R(\Delta t_i - \Delta t')$ is the response function of the \deltat measurement (resolution function), $f_{i}^{\rm sig}$ is the signal probability of the $i$-th event, and $\mathcal{P}_{\rm bkg}$ is the \deltat distribution of background events.
We use a resolution function developed by the Belle collaboration~\cite{ref:BelleResolFunc}.
The resolution function is the convolution of four components: 
detector resolution for the \BCP vertex,
detector resolution for the \Btag vertex,
bias due to secondary particles from charmed intermediate states for the \Btag vertex,
and corrections to the boost factor due to the nonzero c.m.\ momentum of the $B$ mesons. 
The correction to the boost factor is calculated analytically using the cosine of the angle between the \BCP momentum and the boost direction in the $e^+e^-$ c.m.\ frame, \costh, on an event-by-event basis. The resolution-function parameters are fixed to those obtained from a fit to simulated signal events, but the width in simulation is scaled by a parameter \sdet that accounts for data-simulation differences and that is determined simultaneously with \scp and \ccp.
The distribution $\mathcal{P}_{\rm bkg}$ is the sum of two Gaussian functions that depend on vertex quality and vertex-position uncertainty. The $\mathcal{P}_{\rm bkg}$ parameters are determined by a fit to the \mbc sideband data. 
We calculate the signal probability on an event-by-event basis using the five-dimensional PDF of \mbc, \mkkh, \modcs, $r$, and \costh. The PDF contains signal and background components, whose fractions are determined by the signal and background yields. No correlation is assumed between the variables.
The last two variables are included to avoid fit biases (0.03 for \scp and 0.02 for \ccp) due to implicitly assuming equal distributions that differ across sample components~\cite{Punzi:2003wze}.
The $r$ distribution for background is obtained from the $\mbc < 5.265\gevcc$ sideband.
For \costh, we assume a uniform distribution for background and $\frac{3}{4}(1-\costhsq)$ for signal. 
For TI events, we use the likelihood in Eq.~\eqref{eqn_dt4} integrated over \deltat,
\begin{linenomath}
\begin{align} \label{eqn_dt5}
\mathcal{L}^{\rm TI}_i(\ccp|\qfi) = f^{\rm sig}_i \mathcal{P}^{\rm TI}_{\rm sig} (\qfi) + \frac{1-f^{\rm sig}_{i}}{2}.
\end{align}
\end{linenomath}
We include the \ksksk decays in the fit using the likelihood in Eq.~\eqref{eqn_dt4} summed over \qf and using the \Bu lifetime instead of the \Bz lifetime. The control channel helps to constrain \sdet since its signal yield is 2.5 times larger than the TD signal. 
The resolution-function parameters and \sdet are the same as those of the \ksksks events except for the parameters that model the effect of secondary particles. 
They differ since, compared to \Bz mesons, \Bu mesons yield fewer $D^-$ mesons and more $\overbar{D}^0$ mesons, which have shorter lifetimes.
We define the background \deltat distribution for \ksksk with an independent parameter set from \ksksks and with an additional Gaussian function.

\begin{figure}[htb]
  \centering
  \includegraphics[width=0.9\columnwidth]{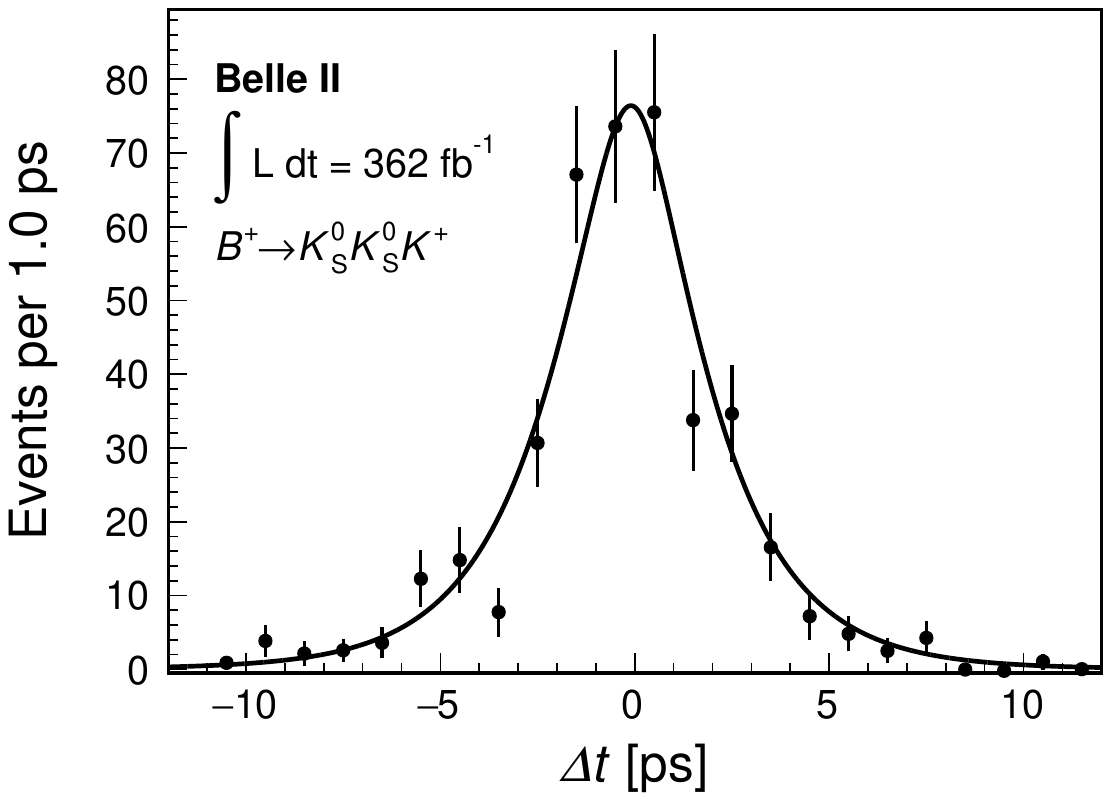}
  \includegraphics[width=0.9\columnwidth]{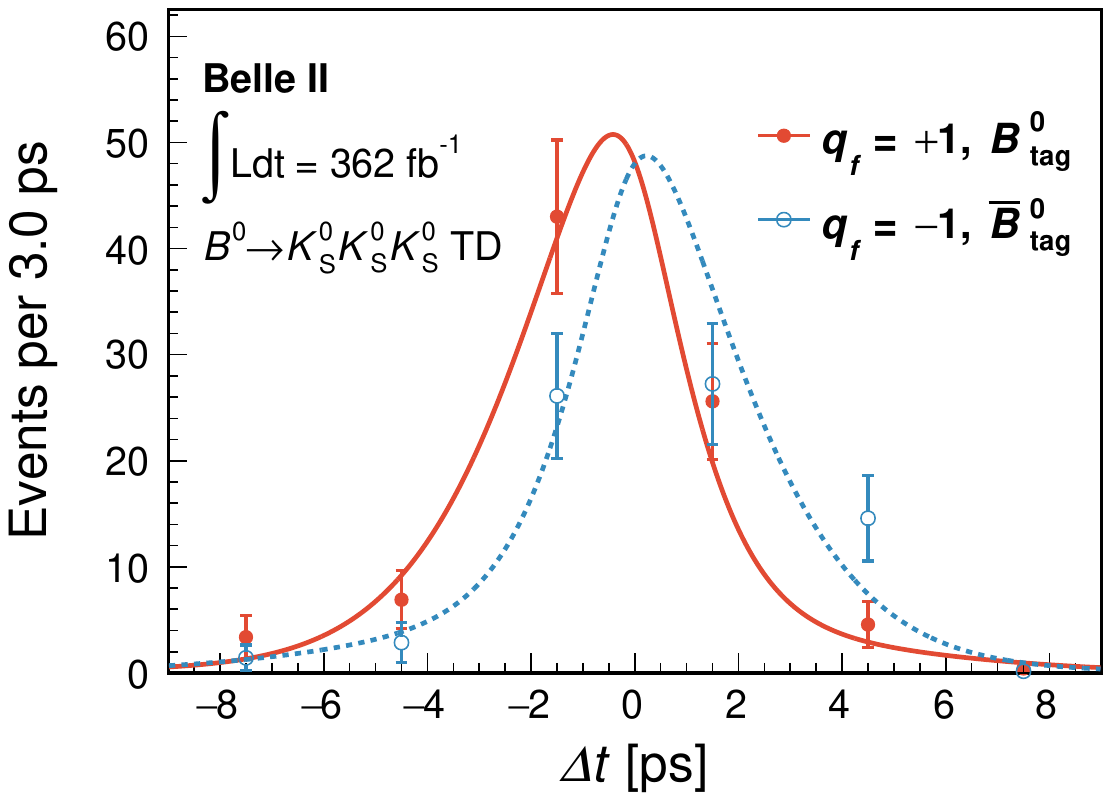}
  \includegraphics[width=0.9\columnwidth]{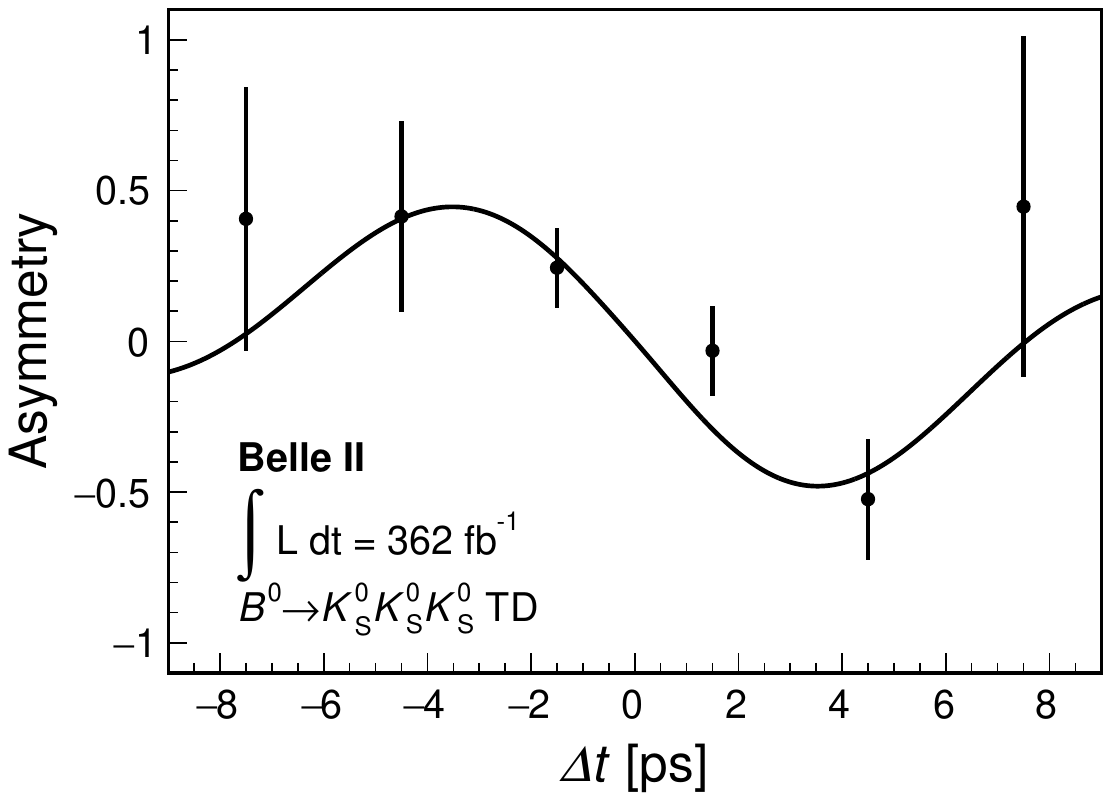}
  \caption{Background-subtracted \deltat distributions for (top) \ksksk candidates  and (middle) \ksksks TD candidates separated for $\qf=\pm1$ along with (bottom) the resulting $B^0_{\rm tag}$ minus $\overline{B}^0_{\rm tag}$ yield-asymmetry as a function of \deltat.  Points with error bars represent data and the curves show the fit results. Red, filled circles and solid curves show the data for $\qf=+1$ and fit results, respectively, while blue, open circles and dashed curves are for $\qf=-1$.}
\label{fig:cpfit}
\end{figure}

Figure~\ref{fig:cpfit} shows the background-subtracted \deltat distributions using the \sPlot technique~\cite{Pivk:2004ty} and their asymmetry with fit projections overlaid.
We obtain $\scp = -1.37_{-0.26}^{+0.32}$, $\ccp = -0.07\pm0.17$, and $\sdet = 1.16 \pm 0.15$. Linear correlation coefficients are $-0.02$ between \scp and \ccp, $-0.16$ between \scp and \sdet, and $-0.07$ between \ccp and \sdet. However, simulation studies show that the above point estimates are not reliable. While the likelihood has no secondary maxima, the small sample size leads to biases and non-Gaussian uncertainties.
For more reliable results, we construct confidence intervals for the \cp-violating parameters as described in Sec.~\ref{sec:results}.

\section{Systematic uncertainties}
\begin{table}[htb]
\caption{Systematic uncertainties}
\label{tab_sys}
\centering
\begin{tabular}
{@{\hspace{0.5cm}}l@{\hspace{0.5cm}}@{\hspace{0.5cm}}c@{\hspace{0.5cm}}@{\hspace{0.5cm}}c@{\hspace{0.5cm}}}
\hline \hline
Source                         & \scp   & \ccp       \\\hline
\tauBz, \tauBu, and $\Delta m_d$ & 0.009  &  0.000    \\ 
Signal modeling                & 0.014  &  0.008    \\ 
\deltat resolution function    & 0.013  &  0.008    \\ 
Background $\Delta t$ modeling & 0.004  &  0.002    \\
Flavor tagging                 & 0.013  &  0.012    \\ 
Fit bias                       & 0.014  &  0.004    \\ 
Tag-side interference          & 0.011  &  0.006    \\
Vertex reconstruction          & 0.011  &  0.004    \\ 
Tracker misalignment           & 0.008  &  0.007    \\\hline 
Total                          & 0.032  &  0.020    \\\hline \hline
\end{tabular}
\end{table}

We consider various sources of systematic uncertainties, which are listed in Table~\ref{tab_sys}.
To evaluate the systematic uncertainties in \scp and \ccp related to assumptions made on parameters of the fit model,
we repeat the fit on data using alternative values of the parameters sampled from Gaussian distributions based on their uncertainties. The widths of the resulting distributions of \scp and \ccp are taken as contributions to the systematic uncertainty.
This approach is used for \tauBz, \tauBu, and $\Delta m_d$; the parameters of the \mbc, \mkkh, and \modcs shapes (referred to as signal modeling in the table); the parameters describing the resolution function; the parameters for the background \deltat shape; and the parameters related to flavor tagging.

We sample the world averages of the \Bz and \Bu lifetimes and $\Delta m_d$ including their uncertainties~\cite{Workman:2022ynf}. The parameters of signal probability, resolution function, and background \deltat shape have uncertainties from the fits used to determine them, which depend on the size of data and simulated samples.
The systematic uncertainty in the resolution function includes the uncertainty due to the choice of the model, which is determined by analyzing a simulated sample with alternative resolution models whose dependence on the vertex-fit quality is partly or entirely removed. The simulation assumes $\scp=-0.7$ and $\ccp=0$.
The systematic uncertainty due to flavor tagging includes the bias due to the flavor asymmetry in the tagging efficiency between \Bz and \Bzb.
Two sets of simplified simulated experiments are generated, with and without the asymmetry, and fits for \scp and \ccp are performed in both ignoring the asymmetry. The difference between the mean values of \scp and \ccp obtained in the two sets is the uncertainty. 
We repeat the simplified simulation assuming various input \cp asymmetries and take the maximum difference.
We observe correlations between \mkkh and vertex-fit quality for \BCP ($-0.06$ for TD events), and between \modcs and $r$ (0.15), which are not included in the default model. 
To evaluate the bias due to these correlations, and to a mismodeling of the \costh distribution, we use simplified simulated samples generated with and without these effects in the same way as above.
The \cp asymmetries are affected by the interference between a CKM-favored transition $\overbar{b} \ra \overbar{c}u\overbar{d}$ and a doubly CKM-suppressed transition $b \ra u \overbar{c}d$ on the tag side~\cite{ref:TSI}. 
We assign as a systematic uncertainty the effect of the tag-side interference assuming the world average values $\scp = -0.83$ and $\ccp = -0.15$\cite{PhysRevD.107.052008}. 
The systematic uncertainty due to the vertex reconstruction is determined by varying the parameters describing the IP profile and boost vector, the track requirements used in the \Btag vertex reconstruction, and the criteria to select TD events, and repeating the fit on data. 
To evaluate the effect from possible misalignment of the vertex detector, we use four simulated samples, each assuming a different misalignment configuration and \cp asymmetries of $\scp = -1.0$ and $\ccp = 0$. We compare the resulting \cp asymmetries with those in the sample without misalignment and the maximum deviation is taken as the systematic uncertainty.

\section{Results and summary}\label{sec:results}
\begin{figure}[htb]
  \centering
  \includegraphics[width=0.9\columnwidth]{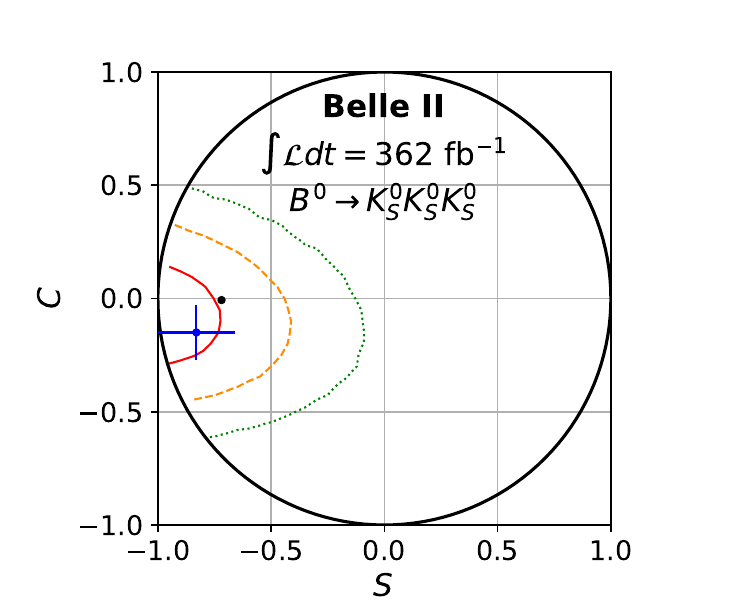}
  \caption{Two-dimensional confidence intervals for \scp and \ccp based on likelihood-ratio ordering. The red solid, orange dashed, and green dotted contours represent the 68.27\%, 95.45\%, and 99.73\% confidence intervals for \scp and \ccp given the physical constraint $\scp^2 + \ccp^2 \leq 1$.
  The blue dot with the error bar is the average value based on results by Belle and BaBar~\cite{Belle:2020cio,BaBar:2011ktx}.  
  The black dot represents the SM prediction $(\scp,\ccp)=(-\sin2\phi_1-0.02,-0.007)$ based on measurements in $B^0\ra (c\overline{c}) K^0$ decays~\cite{PhysRevD.107.052008}.}
\label{fig:contour}
\end{figure}

Since the point estimates from the fit are not reliable, we construct confidence intervals for our results based on likelihood-ratio ordering~\cite{Feldman:1997qc}.
For the construction, simplified simulated experiments are generated by sampling the likelihoods of the yield fit and asymmetry fit. The nuisance parameters in the models are fixed to the values fitted to the data and the systematic uncertainty is not taken into account as its size is negligible.
Figure~\ref{fig:contour} shows the resulting two-dimensional confidence intervals where \scp and \ccp are constrained within their physical boundary, $\scp^2 + \ccp^2 \leq 1$. 
The two-dimensional intervals are $-1 < S < -0.72$ and $-0.29 < C < 0.14$ at the 68.3\% confidence level, $-1 < S < -0.41$ and $-0.45 < C < 0.32$ at the 95.5\% confidence level, and $-1 < S < -0.09$ and $-0.61 < C < 0.49$ at the 99.7\% confidence level.
The results are consistent with the SM predictions and current best determinations by the Belle and BaBar experiments~\cite{Belle:2020cio,BaBar:2011ktx,PhysRevD.107.052008}.

In summary, we report a measurement of decay-time dependent \cp asymmetries in \ksksks decays using a data set of $387 \times 10^6$ \BB pairs reconstructed from electron-positron collisions at the \FourS and collected with Belle II experiment from 2019 to 2022. We reconstruct 220 signal events and extract the \cp-violating parameters from a fit to the distribution of the decay-time difference of the two $B$ mesons. We determine a two-dimensional confidence region for the relevant parameters \scp and \ccp obtaining results that are consistent with the SM predictions and previous determinations.

This work, based on data collected using the Belle II detector, which was built and commissioned prior to March 2019, was supported by
Higher Education and Science Committee of the Republic of Armenia Grant No.~23LCG-1C011;
Australian Research Council and Research Grants
No.~DP200101792, 
No.~DP210101900, 
No.~DP210102831, 
No.~DE220100462, 
No.~LE210100098, 
and
No.~LE230100085; 
Austrian Federal Ministry of Education, Science and Research,
Austrian Science Fund
No.~P~31361-N36
and
No.~J4625-N,
and
Horizon 2020 ERC Starting Grant No.~947006 ``InterLeptons'';
Natural Sciences and Engineering Research Council of Canada, Compute Canada and CANARIE;
National Key R\&D Program of China under Contract No.~2022YFA1601903,
National Natural Science Foundation of China and Research Grants
No.~11575017,
No.~11761141009,
No.~11705209,
No.~11975076,
No.~12135005,
No.~12150004,
No.~12161141008,
and
No.~12175041,
and Shandong Provincial Natural Science Foundation Project~ZR2022JQ02;
the Czech Science Foundation Grant No.~22-18469S;
European Research Council, Seventh Framework PIEF-GA-2013-622527,
Horizon 2020 ERC-Advanced Grants No.~267104 and No.~884719,
Horizon 2020 ERC-Consolidator Grant No.~819127,
Horizon 2020 Marie Sklodowska-Curie Grant Agreement No.~700525 ``NIOBE''
and
No.~101026516,
and
Horizon 2020 Marie Sklodowska-Curie RISE project JENNIFER2 Grant Agreement No.~822070 (European grants);
L'Institut National de Physique Nucl\'{e}aire et de Physique des Particules (IN2P3) du CNRS
and
L'Agence Nationale de la Recherche (ANR) under grant ANR-21-CE31-0009 (France);
BMBF, DFG, HGF, MPG, and AvH Foundation (Germany);
Department of Atomic Energy under Project Identification No.~RTI 4002,
Department of Science and Technology,
and
UPES SEED funding programs
No.~UPES/R\&D-SEED-INFRA/17052023/01 and
No.~UPES/R\&D-SOE/20062022/06 (India);
Israel Science Foundation Grant No.~2476/17,
U.S.-Israel Binational Science Foundation Grant No.~2016113, and
Israel Ministry of Science Grant No.~3-16543;
Istituto Nazionale di Fisica Nucleare and the Research Grants BELLE2;
Japan Society for the Promotion of Science, Grant-in-Aid for Scientific Research Grants
No.~16H03968,
No.~16H03993,
No.~16H06492,
No.~16K05323,
No.~17H01133,
No.~17H05405,
No.~18K03621,
No.~18H03710,
No.~18H05226,
No.~19H00682, 
No.~20H05850,
No.~20H05858,
No.~22H00144,
No.~22K14056,
No.~22K21347,
No.~23H05433,
No.~26220706,
and
No.~26400255,
the National Institute of Informatics, and Science Information NETwork 5 (SINET5), 
and
the Ministry of Education, Culture, Sports, Science, and Technology (MEXT) of Japan;  
National Research Foundation (NRF) of Korea Grants
No.~2016R1\-D1A1B\-02012900,
No.~2018R1\-A2B\-3003643,
No.~2018R1\-A6A1A\-06024970,
No.~2019R1\-I1A3A\-01058933,
No.~2021R1\-A6A1A\-03043957,
No.~2021R1\-F1A\-1060423,
No.~2021R1\-F1A\-1064008,
No.~2022R1\-A2C\-1003993,
and
No.~RS-2022-00197659,
Radiation Science Research Institute,
Foreign Large-Size Research Facility Application Supporting project,
the Global Science Experimental Data Hub Center of the Korea Institute of Science and Technology Information
and
KREONET/GLORIAD;
Universiti Malaya RU grant, Akademi Sains Malaysia, and Ministry of Education Malaysia;
Frontiers of Science Program Contracts
No.~FOINS-296,
No.~CB-221329,
No.~CB-236394,
No.~CB-254409,
and
No.~CB-180023, and SEP-CINVESTAV Research Grant No.~237 (Mexico);
the Polish Ministry of Science and Higher Education and the National Science Center;
the Ministry of Science and Higher Education of the Russian Federation
and
the HSE University Basic Research Program, Moscow;
University of Tabuk Research Grants
No.~S-0256-1438 and No.~S-0280-1439 (Saudi Arabia);
Slovenian Research Agency and Research Grants
No.~J1-9124
and
No.~P1-0135;
Agencia Estatal de Investigacion, Spain
Grant No.~RYC2020-029875-I
and
Generalitat Valenciana, Spain
Grant No.~CIDEGENT/2018/020;
National Science and Technology Council,
and
Ministry of Education (Taiwan);
Thailand Center of Excellence in Physics;
TUBITAK ULAKBIM (Turkey);
National Research Foundation of Ukraine, Project No.~2020.02/0257,
and
Ministry of Education and Science of Ukraine;
the U.S. National Science Foundation and Research Grants
No.~PHY-1913789 
and
No.~PHY-2111604, 
and the U.S. Department of Energy and Research Awards
No.~DE-AC06-76RLO1830, 
No.~DE-SC0007983, 
No.~DE-SC0009824, 
No.~DE-SC0009973, 
No.~DE-SC0010007, 
No.~DE-SC0010073, 
No.~DE-SC0010118, 
No.~DE-SC0010504, 
No.~DE-SC0011784, 
No.~DE-SC0012704, 
No.~DE-SC0019230, 
No.~DE-SC0021274, 
No.~DE-SC0021616, 
No.~DE-SC0022350, 
No.~DE-SC0023470; 
and
the Vietnam Academy of Science and Technology (VAST) under Grants
No.~NVCC.05.12/22-23
and
No.~DL0000.02/24-25.

These acknowledgements are not to be interpreted as an endorsement of any statement made
by any of our institutes, funding agencies, governments, or their representatives.

We thank the SuperKEKB team for delivering high-luminosity collisions;
the KEK cryogenics group for the efficient operation of the detector solenoid magnet;
the KEK computer group and the NII for on-site computing support and SINET6 network support;
and the raw-data centers at BNL, DESY, GridKa, IN2P3, INFN, and the University of Victoria for off-site computing support.

\bibliography{references}

\end{document}